\documentclass[12pt, a4paper, twoside]{article}
\usepackage{graphicx}
\usepackage{color}
\usepackage{lscape}
\usepackage{bm}
\newcommand*{\B}[1]{\ifmmode\bm{#1}\else\textbf{#1}\fi}
\usepackage{epstopdf}
\usepackage{lscape}

\newenvironment{changemargin}[2]{%
\begin{list}{}{%
\setlength{\leftmargin}{#1}%
\setlength{\rightmargin}{#2}%
}%
\item[]}
{\end{list}}
\usepackage{graphicx}
\usepackage{color}

\usepackage{epstopdf}
\usepackage{pdflscape}
\usepackage{amsmath}
\usepackage{amsmath}
\usepackage{amsmath}
\usepackage{kantlipsum}
\allowdisplaybreaks
\usepackage{setspace}

\usepackage{array}

\newcolumntype{M}[1]{>{\raggedright\arraybackslash}m{#1}}
\newcolumntype{N}{@{}m{0pt}@{}}
\usepackage{amsmath}
\usepackage{setspace}
\begin{document}
\baselineskip=0.27in
{\bf \LARGE
\begin{changemargin}{-0.5cm}{-0.5cm}
\begin{center}
{Hydrogen atom in a Laser-Plasma}\footnote{We refer to interaction of plasma with laser light as laser-plasma.}
\end{center}
\end{changemargin}}
\vspace{4mm}
\begin{center}
\large{\bf Babatunde James Falaye $^{a, b, }$}\footnote{\scriptsize E-mail:~ fbjames11@physicist.net;~ babatunde.falaye@fulafia.edu.ng}, {\large{\bf Guo-Hua Sun $^{c, }$}\footnote{\scriptsize E-mail:~ sunghdb@yahoo.com}}, {\large{\bf Muhammed S. Liman $^{b, }$}}\footnote{\scriptsize E-mail:~ limanms2@gmail.com}, {\large{\bf K. J. Oyewumi $^{d, }$}}\footnote{\scriptsize E-mail:~ kjoyewumi66@unilorin.edu.ng},  \large{\bf and} {\large{\bf Shi-Hai Dong $^{e,}$}}\footnote{\scriptsize E-mail:~ dongsh2@yahoo.com }
\end{center}
{\footnotesize
\begin{center}
{\it $^\textbf{a}$Departamento de F\'isica, Escuela Superior de F\'isica y Matem\'aticas, Instituto Polit\'ecnico Nacional, Edificio 9, Unidad Profesional Adolfo L\'opez Mateos, Mexico D.F. 07738, Mexico.} {\it $^\textbf{b}$Applied Theoretical Physics Division, Department of Physics, Federal University Lafia,  P. M. B. 146, Lafia, Nigeria.} {\it $^\textbf{c}$Catedr\'atica CONACyT, CIC, Instituto Polit\'{e}cnico Nacional, Unidad Profesional Adolfo L\'opez Mateos, Mexico D. F. 07700, Mexico.} {\it $^\textbf{d}$Theoretical Physics Section, Department of Physics, University of Ilorin,  P. M. B. 1515, Ilorin, Nigeria.} {\it $^\textbf{e}$CIDETEC, Instituto Polit\'{e}cnico Nacional, Unidad Profesional Adolfo L\'opez Mateos, M\'{e}xico D. F. 07700, M\'{e}xico.}
\end{center}}
\begin{center}
\textit{Laser Phys. Lett. 13 (2016) 116003 (8pp)}
\end{center}
\begin{abstract} 
\noindent
We scrutinize the behaviour of hydrogen atom's eigenvalues in a quantum plasma as it interacts with electric field directed along $\theta=\pi$ and exposed to linearly polarized intense laser field radiation. Using the Kramers-Henneberger (KH) unitary transformation, which is semiclassical counterpart of the Block-Nordsieck transformation in the quantized field formalism, the squared vector potential that appears in the equation of motion is eliminated and the resultant equation is expressed in KH frame. Within this frame, the resulting potential and the corresponding wavefunction have been expanded in Fourier series and using Ehlotzky's approximation, we obtain a laser-dressed potential to simulate intense laser field. By fitting the exponential-cosine-screened Coulomb potential into the laser-dressed potential, and then expanding it in Taylor series up to $\mathcal{O}(r^4,\alpha_0^9)$, we obtain the eigensolution (eigenvalues and wavefunction) of hydrogen atom in laser-plasma encircled by electric field, within the framework of perturbation theory formalism. Our numerical results show that for a weak external electric field and gargantuan length of Debye screening parameter, the system is strongly repulsive in contrast for strong external electric field and small length of Debye screening parameter, the system is very attractive. This work has potential application in the areas of atomic and molecular processes in external fields including interactions with strong fields and short pulses.
\noindent
\end{abstract}


Hydrogen atom.

{\bf PACs No.}: 34.50.Fa, 52.25.Kn, 52.27.Gr, 52.27.Lw

\section{Introduction}
Lasers have emerged as one of the world's indispensable technologies, employed in telecommunications, law enforcement, military equipment, etc. Recent furtherance in laser technology has aroused the interest of many researchers to investigate new sources of laser in order to probe and control molecular structure, function and dynamics on the natural timescale of atomic motion, the femtosecond and electron motion on attosecond timescale \cite{BJ1}. To obtain intense laser fields, it is required to concentrate large amounts of energy within short period of time, and then focus the laser light onto a small area. In an intense laser system, a train of pulses of short duration are created by the oscillator. The energy of the pulses is then proliferated by the amplifier, which is eventually focused.

Studying atoms in intense laser fields have been a subject of active research for more than three decades due to its salient application in the invention of high-power short-pulse laser technologies. These atoms exhibit new properties that have been discovered via the study of multiphoton processes. When a high-power laser is directed into a gas of atoms, the magnitude of electromagnetic field is found to be consistent with the Coulomb field, which binds a 1s electron in a Hydrogen atom \cite{BJ2}. Within this context, so many outstanding results have been reported so far (see \cite{BJ3,BJ4,BJ5,BJ6, BJ7} and refs. therein). It was shown in ref.  \cite{BJ8}, that in the presence of an oscillating magnetic field, the ionization rate due to the laser field  dwindle, and the electron density becomes ionized with a less rate by keeping the magnetic field strength constant and increasing the intensity of the laser. 

There has been renewable interest (see \cite{BJ11,BJ13,BJ14,BJ15,BJ16} and references therein) in studying atomic and molecular processes in the quantum plasma environment due to their applications in distinguishing various plasmas and also providing passable knowledge of collision dynamics \cite{BJ11}.  The role which ionization processes and atomic excitation play in the conceptual understanding of various phenomena related to hot plasma physics and astrophysics is preeminent. The effects of quantum plasma environment atoms can be modeled by screened potential which accounts for pair correlations. In accord with, this enormous number of studies have been reported to investigate the influence external fields on hydrogen atom in quantum plasma (see \cite{BJ11,BJ13,BJ17} and refs. therein). Very recently, Falaye et al. \cite{BJ11} found that to perpetuate a low-energy medium for the hydrogen atom in quantum plasmas, a strong electric field and weak magnetic field are required, whereas the Aharonov–Bohm flux field can be used as a regulator.

Researchers have recently developed kin interest in scrutinizing atomic processes in laser-plasma. Within this context, Idris et al. \cite{BJ18}, Kurniawana and Kagawab \cite{BJ19} had examined hydrogen emission in laser plasma via focusing  a TEA CO2 laser and Nd-YAG laser on various types of samples doped with hydrogen. Some other outstanding reports can be found in refs. (\cite{BJ20,BJ21,BJ22,BJ23} and refs. therein). However, it is worth mentioning that most of these worthy attempts were experimentally based. In the present work, our objective is to scrutinize the behaviour of hydrogen atom's eigenvalues in a quantum plasma as it interacts with electric field, and exposes to linearly polarized intense laser field radiation. To our best knowledge, this study has not been reported yet and in fact it represents a significant furtherance of refs. \cite{BJ6,BJ11}. Consequently, we feel this work will be of interest in the areas of atomic structure and collisions in plasmas. 

\section{Formulation of the problem}
In this section, we derive the equation of motion for spherically confined hydrogen atom in a dense quantum plasma under electric field, and expose to linearly polarized intense laser field radiation. In order to achieve the goal of this section, we start with the following time-dependent Schr\"odinger wave equation
\begin{equation}
i\hbar\frac{\partial}{\partial t}\Psi(\bm{r},t)=\left[-\frac{\hbar^2}{2\mu}{\nabla}^2-i\hbar\frac{e}{2\mu}\big[\bm{A}(\bm{r},t)\cdot{\nabla}+{\nabla}\cdot {\bf A}(\bm{r},t)\big]+\frac{e^2}{2\mu}{\bf A}(\bm{r},t)^2-e\phi+V(\bm{r})+Fr\right]\Psi(\bm{r},t),
\label{EQ1}
\end{equation}
with the scalar potential $\phi({\bf r},t)$ and the vector potential $\bm{A}(\bm{r},t)$ which is invariant under the gauge transformation. $\mu$ is the effective mass of the electron. Furthermore, $F$ denotes an electric field strength with an angle $\theta$ between $F$ and $r$. With $\theta = \pi$, $-Fr\cos(\theta)$ becomes $Fr$ as shown in Eq. (\ref{EQ1}). We consider Coulomb gauge, such that $\nabla\cdot {\bf A}(\bm{r},t)=0$  with $\phi=0$ in empty space and then simplify the interaction term in the equation (\ref{EQ1}) by performing gauge transformations within the framework of dipole approximation. In this approximation, for an atom whose nucleus is located at the position $r_0$, the vector potential is spatially homogeneous ${\bf A}(\bm{r},t)\approx\bm{A}(t)$. Moreover, term ${\bf A}({\bf r},t)^2$ appearing in equation (\ref{EQ1}) is considered for extremely high field strength. It is usually small and can be eliminated by extracting a time-dependent phase factor from the wave function via \cite{BJ24}
\begin{equation}
\Psi^{v}(\bm{r},t)=\exp\left[\frac{ie^2}{2\mu\hbar}\int_{-\infty}^t{\bf A}(t')^2dt'\right]\Psi(\bm{r},t),
\label{EQ2}
\end{equation}
to obtain {velocity gauge}\footnote{Because the vector potential A(t) is being coupled to the operator $\bm{p}/m$ via the interaction Hamiltonian. $\bm{p}=-ih{\nabla}.$}
\begin{equation}
i\hbar\frac{\partial}{\partial t}\Psi^{v}({\bf r},t)=\left[-\frac{\hbar^2}{2\mu}\nabla^2-i\hbar\frac{e}{\mu}{\bf A}(t)\cdot\nabla+V(r)+Fr\right]\Psi^{v}({\bf r},t).
\label{EQ3}
\end{equation}
A prerequisite to study hydrogen atom in intense high-frequency laser field is transforming equation (\ref{EQ3}) to the Kramers-Henneberger accelerated frame. Now, with the introduction of the following unitary Kramers-Henneberger's transformation
\begin{equation}
\Psi^{A}({\bf r},t)= U^\dag\Psi^{v}({\bf r},t)\ \mbox{with} \ \ \ U=\exp\left[-\frac{i}{\hbar}\bm{\alpha}(t).\bm{p}\right],\ \ \mbox{and} \ \bm{\alpha}(t)=\frac{e}{\mu}\int^t{\bf A}(t')dt',
\label{EQ4}
\end{equation}
which is semiclassical counterpart of the Block-Nordsieck transformation in the quantized field formalism, the coupling term $\bm{A}(t)\cdot\bm{p}$ in the velocity guage (i.e., Eq. (\ref{EQ3})) is eliminated. More explicitly, this can be done via
\begin{equation}
i\hbar U^\dag\frac{\partial}{\partial t}U\Psi^{A}(\bm{r},t)=U^\dag\left[-\frac{\hbar^2}{2\mu}\nabla^2-i\hbar\frac{e}{\mu}{\bf A}(t)\cdot{\nabla}+V(\bm{r})+Fr\right]U\Psi^{A}({\bf r},t).
\label{EQ5}
\end{equation}
Evaluation of terms in equation (\ref{EQ5}) are straightforward and easy. However, let us try to be more explicit in evaluating the term $U^\dag V(r)U$. This can be done via Campbell-Baker-Hausdorff identity: $e^{\hat{A}}\hat{B}e^{-\hat{A}}=\hat{B}+[\hat{A},\hat{B}]+[\hat{A},[\hat{A},\hat{B}]]/2!+\ldots$. Thus, we have
\begin{eqnarray}
U^\dag V(r)U&=&\exp\left[\frac{i}{\hbar}\bm{\alpha}(t).\bm{p}\right]V(\bm{r})\exp\left[-\frac{i}{\hbar}\bm{\alpha}(t).\bm{p}\right]\nonumber\\
&=&V(\bm{r})+\left[\bm{\alpha}(t).\nabla\right]V(\bm{r})+\frac{1}{2!}\left[\bm{\alpha}(t).\nabla\right]^2V(\bm{r})+\ldots\nonumber\\
&=&V\left[r+\bm{\alpha}(t)\right],
\label{EQ6}
\end{eqnarray}
where $\bm{\alpha}(t)$ denotes the displacement of a free electron in the incident laser field. Ergo, Eq. (\ref{EQ5}) becomes
\begin{equation}
i\hbar\frac{\partial}{\partial t}\Psi^{A}(\bm{r},t)= -\frac{\hbar^2}{2\mu}\nabla^2\Psi^{A}(\bm{r},t)+V\left[\bm{r}+\bm{\alpha}(t)\right]\Psi^{A}(\bm{r},t)+Fr\Psi^{A}(\bm{r},t).
\label{EQ7}
\end{equation}
Eq. (\ref{EQ7}) represents a space-translated version of the time-dependent Schr\"odinger wave equation with incorporation of $\bm{\alpha}(t)$ into the potential in order to simulate the interaction of atomic system with the laser field. Three decades after its discovery by Pauli and Fierz \cite{BJ25}, it was applied to study renormalization of quantum electrodynamics by Kramers \cite {BJ26} and was later used to study interactions of atoms with laser by Henneberger \cite{BJ27}. Within this framework, so many outstanding works have been reported by great number of erudite scholars (see \cite{BJ4, BJ28, BJ29,BJ30}, and refs. therein).

For a steady field condition, the vector potential takes the form $\bm{A}(t)=(\mathcal{E}_0/\omega)\cos(\omega t)$ with $\bm{\alpha}(t)=\alpha_0\sin{(\omega t)}$, where $\alpha_0=e\mathcal{E}_0/(\mu\omega^2)$ is the amplitude of oscillation of a free electron in the field (called as {\it laser-dressing parameter}), $\mathcal{E}_0$ denotes the amplitude of electromagnetic field strength and $\omega$ is the angular frequency. Now, considering a pulse where the electric field amplitude is steady, the wavefunction in the frame of Kramers-Henneberger takes the following Floquet form \cite{BJ24}:
\begin{equation}
\Psi^{A}(\bm{r},t)=e^{-\frac{iE_{_{KH}}}{\hbar}t}\sum_{n}\Psi_n^{E_{_{KH}}}(\bm{r})e^{-in\omega t},
\label{EQ8}
\end{equation}
where Floquet quasi-energy has been denoted by $E_{KH}$. The potential in the frame of Kramers-Henneberger can be expanded in Fourier series as \cite{BJ29}
\begin{equation}
V[\bm{r}+\bm{\alpha}(t)]=\sum_{m=-\infty}^\infty V_m(\alpha_0;\bm{r})e^{-im\omega t}\ \ \ \mbox{with}\ \ \ V_m(\alpha_0;\bm{r})=\frac{i^m}{\pi}\int_{-1}^{1} V(\bm{r}+{\alpha}_0\varrho)\frac{T_n(\varrho)}{\sqrt{1-\varrho^2}}d\varrho,
\label{EQ9}
\end{equation}
where we have taken the period as $2\pi/\omega$ and introduced a new transformation of the form $\varrho=\sin(\omega t)$. Furthermore, $T_n(\varrho)$ are Chebyshev polynomials. Substituting Eqs. (\ref{EQ8}) and (\ref{EQ9}) into Eq. (\ref{EQ7}), yields a set of coupled differential equation:
\begin{equation}
\left[-\frac{\hbar^2}{2\mu}\nabla^2+V_m(\alpha_0;\bm{r})+Fr-(E_{KH}+n\hbar\omega)\right]\Psi_n^{E_{_{KH}}}(\bm{r})=-\sum_{m=-\infty}^{\infty, m\neq n} V_{n-m}\Psi_m^{E_{_{KH}}}(\bm{r}).
\label{EQ10}
\end{equation}
Considering $n=0$ (which gives the lowest order approximation) and high frequency limit (which made $V_{m}$ with $m\neq0$ vanish), Eq. (\ref{EQ10}) becomes
\begin{equation}
\left[-\frac{\hbar^2}{2\mu}\nabla^2+V_0(\alpha_0;\bm{r})+Fr-E_{KH}\right]\Psi_0^{E_{_{KH}}}(\bm{r})=0.
\label{EQ11}
\end{equation}
and the coefficient of the Fourier series for the potential becomes
\begin{eqnarray}
V_0(\alpha_0;\bm{r})&=&\frac{1}{\pi}\int_{-1}^{1}V(\bm{r}+\alpha_0\varrho)\frac{d\varrho}{\sqrt{1-\varrho^2}}
=\frac{1}{\pi}\int_{0}^{1}\left[V(\bm{r}+\alpha_0\varrho)+V(\bm{r}-\alpha_0\varrho)\right]\frac{d\varrho}{\sqrt{1-\varrho^2}}.
\label{EQ12}
\end{eqnarray}
Using Ehlotzky approximation \cite{BJ31}, one has $[V(\bm{r}+\alpha_0\varrho)+V(\bm{r}-\alpha_0\varrho)]\approx [V(\bm{r}+\alpha_0)+V(\bm{r}-\alpha_0)]$. Hence, by evaluating the integral, we obtain
\begin{eqnarray}
V_0(\alpha_0;\bm{r})=\frac{1}{2}\left[V(\bm{r}+\alpha_0)+V(\bm{r}-\alpha_0)\right].
\label{EQ13}
\end{eqnarray}
Eq. (\ref{EQ13}) is the approximate expression to model laser field. Now, we incorporate the model to simulate the behaviour of hydrogen atom in dense quantum plasma \cite{BJ6, BJ11} into model potential (\ref{EQ13}). Then Eq. (\ref{EQ11}) becomes
\begin{equation}
\left[-\frac{\hbar^2}{2\mu}\nabla^2-\frac{Ze^2}{r_{\alpha_0^+}}\exp\left(-\frac{r_{\alpha_0^+}}{\lambda_D}\right)\cos\left(\frac{r_{\alpha_0^+}}{\lambda_D}\right)-\frac{Ze^2}{r_{\alpha_0^-}}\exp\left(-\frac{r_{\alpha_0^-}}{\lambda_D}\right)\cos\left(\frac{r_{\alpha_0^-}}{\lambda_D}\right)+Fr-E_{KH}\right]\Psi_0^{E_{_{KH}}}(\bm{r})=0,
\label{EQ14}
\end{equation}
where $r_{\alpha_0^{\pm}}=r\pm\alpha_0$, $\lambda_D$ is the Debye screening length and $Z$ denotes the atomic number that
is found useful in describing energy levels of light to heavy neutral atoms \cite{BJ11}. We have assumed the core of the hydrogenic system to be static which explains the reason behind one body system appearing in Eq. (\ref{EQ14}) instead of two-body.  The above equation (\ref{EQ14}) is the equation motion for spherically confined hydrogen atom in a dense quantum plasma under electric field, and expose to linearly polarized intense laser field radiation. To achieve our goal in this study, in the next section, we solve Eq. (\ref{EQ14}), using perturbation formalism.
\begin{figure*}[!t]
\centering \includegraphics[height=140mm, width=165mm]{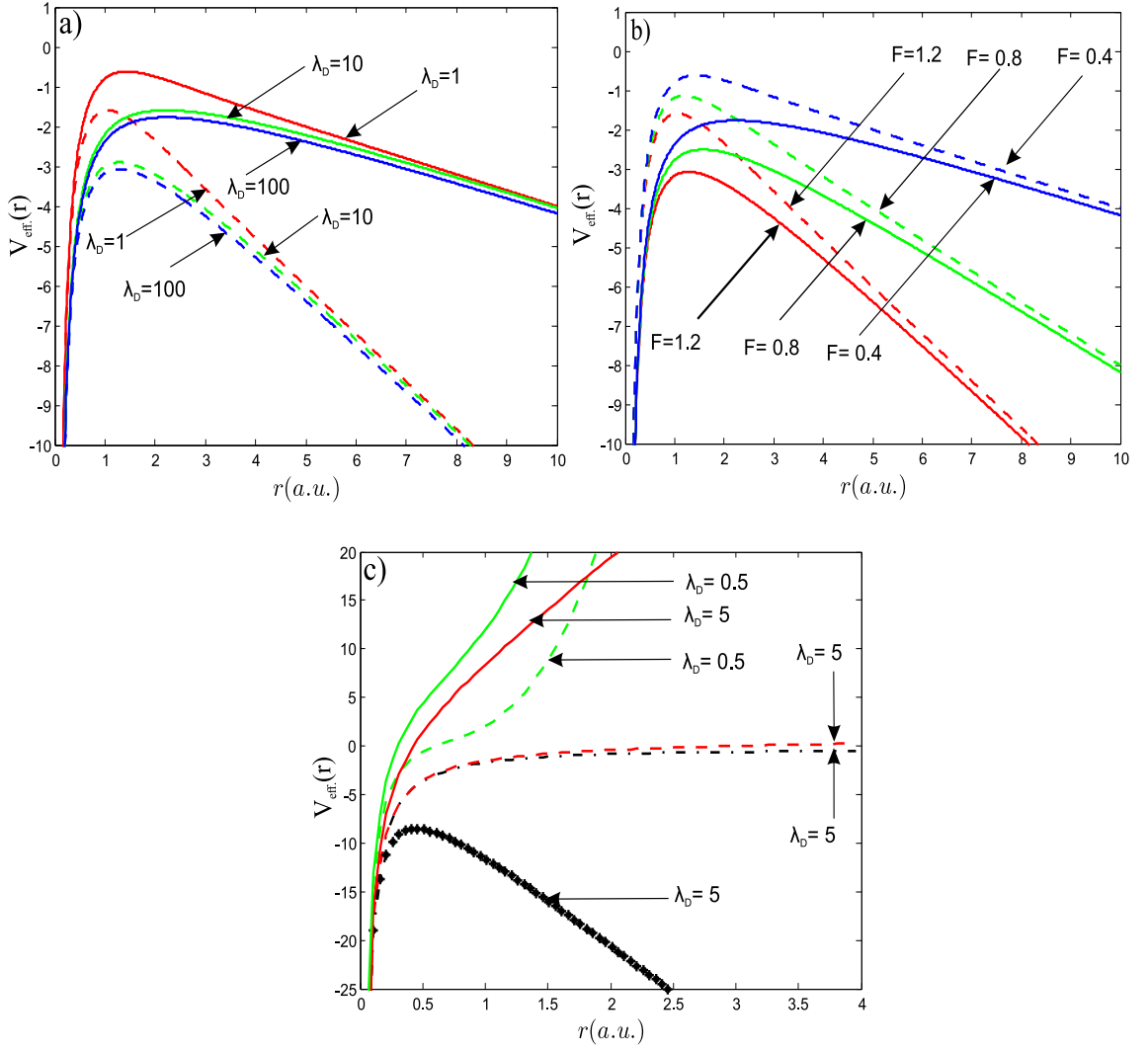}
\caption{\protect\footnotesize The plots of model potential as a function of internuclear distance. In (a), we depict the shape of the model by considering a weak and strong external electric field via taking $F$ as $0.4$ and $1.2$ respectively for various values of Debye screening lengths $\lambda_D$. The dash line ``- -" represents the context of weak $F$ while the thick line represents strong $F$. Furthermore, in (b),  we depict the shape of the model potential by taking Debye length $\lambda_D$ as $1$ (dash lines) and $100$ (thick lines)  for various values of electric field strengths. The figures (a) and (b) show that increasing the electric field strength will reduce the attractiveness of the effective potential and ergo; the system becomes more repulsive. In figure (c), we show the accuracy of the approximate expression for the potential model (i.e., the Taylor series expansion of the effective potential) given in Eq. (\ref{EQ16}) for various values of $\lambda_D$ and $F$. Line with  marker type ``$_*$" represents the effective potential for $F=10$ while the dot-dash line ``. -" is for $F=0.1$. The thick lines represent approximate expression for the effective potential with $F=10$ while the dash line is for $F=0.1$. This figure shows that the approximation is only valid for low $F$ and high $\lambda_D$. It can also be inferred that, for the validity of the approximation to be maintained, $r/\lambda_D$ must be $<<1$. We have taken $\alpha_0=0.001$ and all our numerical computations are in atomic units (a.u.).} 
\label{fig1}
\end{figure*}
\section{Eigenspectra Calculation}
\begin{figure*}[!h]
\centering \includegraphics[height=130mm, width=155mm]{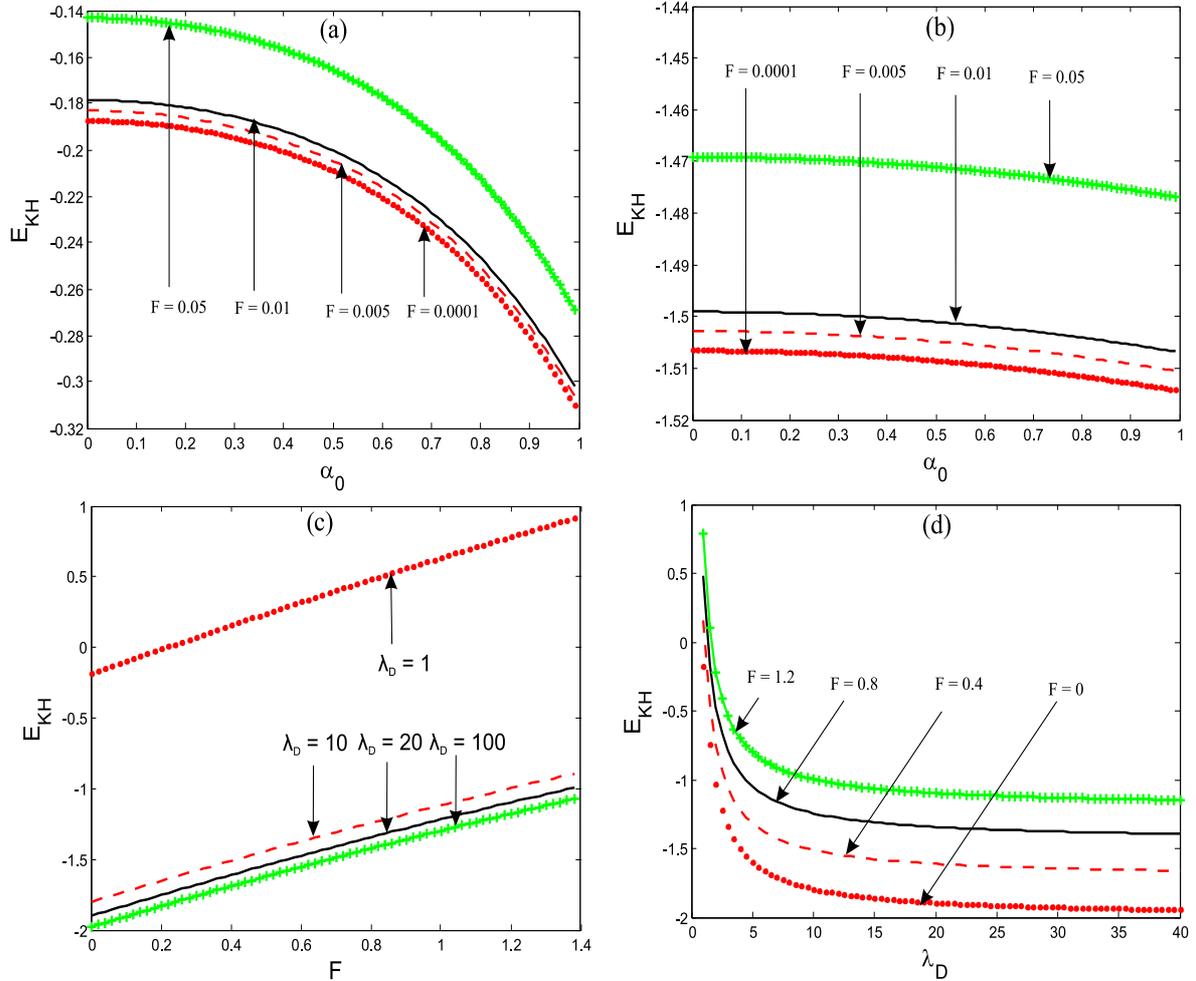}
\caption{\protect\footnotesize Plots of energy eigenvalues of hydrogen atom in quantum plasma as it interacts with electric field and exposed to linearly polarized intense laser field radiation, as a function of various model parameters. (a) Energy eigenvalues as a function of laser-dressing parameter with $\lambda_D=1$ for various values of electric field intensities. (b) Same as (a) but for $\lambda_D=4$. (c) Eigenvalues as a function of electric field with $\alpha_0=0.0001$ and for various values of Debye screening lengths. (d) Eigenvalues as a function of Debye screening length with $\alpha_0=0.0001$ and for various values of electric field intensities.  All our numerical computations are in a.u..} 
\label{fig2}
\end{figure*}

Eq. (\ref{EQ14}) is not solvable analytically. One can either use numerical procedure or perturbation formalism. Using perturbation approach, we decompose the equation into two parts where the first part is exactly solvable and the other part is perturbation. Consequently, the eigenvalue solutions are represented in power series with the leading term corresponding to the solution of exactly solvable part and the other part is correction to the energy term which corresponds to the perturbation term. This approach has been used in numerous research reports (See \cite{BJ11,BJ32} and references therein). Now, we re-write Eq. (\ref{EQ14}) as
\begin{equation}
\frac{\hbar^2}{2\mu}\left(\frac{\nabla^2\mathcal{X}_0(r)}{\mathcal{X}_0(r)}+\frac{\nabla^2\mathcal{Y}_0(r)}{\mathcal{Y}_0(r)}+2\frac{\nabla\mathcal{X}_0(r)\nabla\mathcal{Y}_0(r)}{\mathcal{X}_0(r)\nabla\mathcal{Y}_0(r)}\right)=V_{\rm eff.}(r)-E_{KH},
\label{EQ15}
\end{equation}
where $\Psi_0^{E_{_{KH}}}(\bm{r})=\mathcal{X}_0(r)\mathcal{Y}_0(r)$ with $\mathcal{X}_0(r)$ as the wavefunction of the exactly solvable part and $\mathcal{Y}_0(r)$ as the moderating wavefunction. The effective potential $V_{\rm eff.}(r)$ represents the Taylor's series expansion of the potential terms in Eq. (\ref{EQ14}). This can be written as:
\begin{eqnarray}
V_{\rm eff.}(r)&=&-\frac{2\mathcal{A}}{r}+\left(\frac{\mathcal{A}\alpha_0^8}{11340\lambda_D^9}+\frac{\mathcal{A}\alpha_0^6}{315\lambda_D ^7}-\frac{\mathcal{A}\alpha_0^4}{15\lambda_D ^5}-\frac{2\mathcal{A}\alpha_0^2}{3\lambda_D ^3}+\frac{2\mathcal{A}}{\lambda_D}\right)+r\left(F-\frac{\mathcal{A}\alpha_0^6}{180\lambda_D^8}+\frac{\mathcal{A}\alpha_0^2}{\lambda_D ^4}\right)\nonumber\\
&&+r^2 \left(-\frac{\mathcal{A}\alpha_0^8}{13860 \lambda_D^{11}}+\frac{\mathcal{A}\alpha_0^6}{405\lambda_D^9}+\frac{\mathcal{A}\alpha_0^4}{21\lambda_D^7}-\frac{2\mathcal{A}\alpha_0^2}{5 \lambda_D^5}-\frac{2\mathcal{A}}{3\lambda_D^3}\right)\nonumber\\
&&+r^3 \left(\frac{\mathcal{A}\alpha_0^8}{22680\lambda_D^{12}}-\frac{\mathcal{A}\alpha_0^4}{36\lambda_D^8}+\frac{\mathcal{A}}{3\lambda_D^4}\right)+\mathcal{O}(r^4,\alpha_0^9), \ \ \ \ \mathcal{A}=Ze^2.
\label{EQ16}
\end{eqnarray}
The first term is the main part which corresponds to a shape invariant potential for which the superpotential is known analytically and the remaining part is taken as a perturbation, $\Delta V_{\rm eff.}(r)$. This approximation is only valid for $r/\lambda_D<<1$. The effective potential and its approximate expansion have been delineated in Figure \ref{fig1}. Now, taking the logarithmic derivatives of the perturbed and unperturbed wavefunctions as $W_0(r)=-(\hbar/\sqrt{2\mu})\mathcal{X}_0'/\mathcal{X}_0$ and $\Delta W_0(r)=-(\hbar/\sqrt{2\mu})\mathcal{Y}_0'/\mathcal{Y}_0$, and then substitute them into (\ref{EQ15}), yield the following equation
\begin{subequations} 
\begin{equation}
\frac{\hbar^2}{2\mu}\frac{\mathcal{X}_0''(r)}{\mathcal{X}_0(r)}=W_0^2(r)-\frac{\hbar}{\sqrt{2\mu}}W_0'(r)= -\frac{2\mathcal{A}}{r}-E_{KH}^{(0)},
\label{EQ17a}
\end{equation}
\begin{equation}
\Delta W_0^2(r)-\frac{\hbar}{\sqrt{2\mu}}\Delta W_0'(r)+2W_0(r)\Delta W_0(r)=\Delta V_{\rm eff.}(r)-\Delta E_{KH},
\label{EQ17b}
\end{equation}
\end{subequations} 
where $E_{KH}^{(0)}$ is the eigenvalue of exactly solvable part and $\Delta E_{KH}=E_{KH}^{(1)}+E_{KH}^{(2)}+E_{KH}^{(3)}+...$ is correction to the energy which corresponds to the perturbation term. Eq. (\ref{EQ17a}) is analytically solvable via formula method \cite{BJ33} to obtain
\begin{equation}
\mathcal{X}_0(r)=2\varsigma^{3/2}re^{-\varsigma r},\ \ \ W_0(r)=-\frac{\hbar}{r\sqrt{2\mu}}+\frac{\mathcal{A}\sqrt{2\mu}}{\hbar},\ \ \ E_{KH}^{(0)}=- \varsigma\mathcal{A},\ \ \mbox{where}\ \ \varsigma=\frac{2\mu \mathcal{A}}{\hbar^2}.
\label{EQ18}
\end{equation}
On the contrary, Eq. (\ref{EQ17b}) is not exactly solvable. It is therefore required to expand the related functions as
$\Delta V_{\rm eff.}(r;\eta)=\sum_{i=1}^\infty\eta_iV_{\rm eff.}(r)^{(i)}$, $\Delta W_{0}(r;\eta)=\sum_{i=1}^\infty\eta_iW_{0}^{(i)}$, $\Delta E_{0}^{(i)}(\eta)=\sum_{i=1}^\infty\eta_iE_{0}^{(i)}$, where $i$ represents the order of perturbation. We substitute these expressions into equation (\ref{EQ17b}) and then equate terms with same power of $\eta$ on both sides to have the following expressions
\begin{subequations}
\begin{eqnarray}
2W_{0}(r)W_{0}^{(1)}(r)-\frac{\hbar}{\sqrt{2\mu}}\frac{dW_{0}^{(1)}(r)}{dr}=V_{\rm eff.}^{(1)}(r)-E_{KH}^{(1)}, \label{EQ19a}\\
\left[W_{0}^{(1)}(r)\right]^2+2W_{0}(r)W_{0}^{(2)}(r)-\frac{\hbar}{\sqrt{2\mu}}\frac{dW_{0}^{(2)}(r)}{dr}=V_{\rm eff.}^{(2)}(r)-E_{KH}^{(2)}, \label{EQ19b}\\
2\left[W_{0}(r)W_{0}^{(3)}(r)+W_{0}^{(1)}(r)W_{0}^{(2)}(r)\right]-\frac{\hbar}{\sqrt{2\mu}}\frac{dW_{0}^{(3)}(r)}{dr}=V_{\rm eff.}^{(3)}(r)-E_{KH}^{(3)}, \label{EQ19c}\\
2\left[W_{0}(r)W_{0}^{(4)}(r)+W_{0}^{(1)}(r)W_{0}^{(3)}(r)\right]+W_{0}^{(2)}(r)W_{0}^{(2)}(r)-\frac{\hbar}{\sqrt{2\mu}}\frac{dW_{0}^{(4)}(r)}{dr}=V_{\rm eff.}^{(4)}(r)-E_{KH}^{(4)}.
\label{EQ19d}
\end{eqnarray}
\end{subequations}
Taking the superpotentials into account and then multiplying each term in equations (\ref{EQ19a}-\ref{EQ19d}) by $\mathcal{X}_0^2(r)$, we obtain first, second and third -order corrections to the energy and their superpotentials as follows:
\begin{subequations}
\begin{eqnarray}
E_{KH}^{(1)}&=&\int_{0}^{\infty}\mathcal{X}_{0}^2(r)r\left(F-\frac{\mathcal{A}\alpha_0^6}{180\lambda_D^8}+\frac{\mathcal{A}\alpha_0^2}{\lambda_D ^4}\right)dr=\frac{3}{2\varsigma}\left(F-\frac{\mathcal{A}\alpha_0^6}{180\lambda_D^8}+\frac{\mathcal{A}\alpha_0^2}{\lambda_D ^4}\right),
\label{EQ20a}\\
W_{0}^{(1)}(r)&=&\sqrt{\frac{2\mu}{\hbar^2}}\frac{1}{\mathcal{X}_{0}^2(r)}\int_{}^{r}\mathcal{X}_{0}^2(\varrho)\left[E_{KH}^{(1)}-\left(F-\frac{\mathcal{A}\alpha_0^6}{180\lambda_D^8}+\frac{\mathcal{A}\alpha_0^2}{\lambda_D ^4}\right)\varrho\right]d\varrho\nonumber\\
&=&\frac{r}{\hbar\varsigma}\sqrt{\frac{\mu}{2}}\left(F-\frac{\mathcal{A}\alpha_0^6}{180\lambda_D^8}+\frac{\mathcal{A}\alpha_0^2}{\lambda_D ^4}\right)\label{EQ20b}\\
E_{KH}^{(2)}&=&\int_{0}^{\infty}\mathcal{X}_{0}^2(r)\left[r^2 \left(-\frac{\mathcal{A}\alpha_0^8}{13860 \lambda_D^{11}}+\frac{\mathcal{A}\alpha_0^6}{405\lambda_D^9}+\frac{\mathcal{A}\alpha_0^4}{21\lambda_D^7}-\frac{2\mathcal{A}\alpha_0^2}{5 \lambda_D^5}-\frac{2\mathcal{A}}{3\lambda_D^3}\right)-{W_{0}^{(1)}}^2(r)\right]dr\nonumber\\
&=&\bigg[\frac{\hbar^4}{4\mu^2\mathcal{A}}\left(-\frac{\alpha_0^8}{4620 \lambda_D^{11}}+\frac{\alpha_0^6}{135\lambda_D^9}+\frac{\alpha_0^4}{7\lambda_D^7}-\frac{6\alpha_0^2}{5 \lambda_D^5}-\frac{2}{\lambda_D^3}\right)\nonumber\\
&&\ \ \ \ \ \ \ \ \ \ \ \ \ \ \ \ \ \ \ \ \  \ \ \ \ \ \ \ \ \ \ \ \ \ \ -\frac{3\hbar^6}{32\mu^3\mathcal{A}^2}\left(\frac{F}{\mathcal{A}}-\frac{\alpha_0^6}{180\lambda_D^8}+\frac{\alpha_0^2}{\lambda_D ^4}\right)^2\bigg]\label{EQ20c}\\
W_{0}^{(2)}(r)&=&\sqrt{\frac{2\mu}{\hbar^2}}\frac{1}{\mathcal{X}_{0}^2(r)}\bigg[\int_{}^{r}\mathcal{X}_{0}^2(\varrho)\left(E_{KH}^{(2)}+{W_{0}^{(1)}}^2(\varrho)\right)d\varrho.\nonumber\\
&&\ \ \ \ \ \ \ \ \ \ \ \ \  -\int_{}^{r}\mathcal{X}_{0}^2(\varrho)\left(-\frac{\mathcal{A}\alpha_0^8}{13860 \lambda_D^{11}}+\frac{\mathcal{A}\alpha_0^6}{405\lambda_D^9}+\frac{\mathcal{A}\alpha_0^4}{21\lambda_D^7}-\frac{2\mathcal{A}\alpha_0^2}{5 \lambda_D^5}-\frac{2\mathcal{A}}{3\lambda_D^3}\right)\varrho^2d\varrho\bigg]\nonumber\\
&=&\frac{r}{\varsigma\hbar}\sqrt{\frac{\mu}{2}}\left(r+\frac{2}{\varsigma}\right)\left(\frac{\mathcal{A}\alpha_0^8}{13860 \lambda_D^{11}}+\frac{\mathcal{A}\alpha_0^6}{405\lambda_D^9}+\frac{\mathcal{A}\alpha_0^4}{21\lambda_D^7}-\frac{2\mathcal{A}\alpha_0^2}{5 \lambda_D^5}-\frac{2\mathcal{A}}{3\lambda_D^3}\right)\nonumber\\
&&\ \ \ \ \ \ \ \ \ \ \ \ \ \ \ \ \ \ \ \ \ \ \ \ \ \ -\frac{r}{\varsigma^3\hbar^3}\left(\frac{\mu}{2}\right)^{3/2}\left(r+\frac{2}{\varsigma}\right)\left(F-\frac{\mathcal{A}\alpha_0^6}{180\lambda_D^8}+\frac{\mathcal{A}\alpha_0^2}{\lambda_D ^4}\right)^2\label{EQ20d}\\
E_{KH}^{(3)}&=&\int_{0}^{\infty}\mathcal{X}_{0}^2(r)\left[r^3 \left(\frac{\mathcal{A}\alpha_0^8}{22680\lambda_D^{12}}-\frac{\mathcal{A}\alpha_0^4}{36\lambda_D^8}+\frac{\mathcal{A}}{3\lambda_D^4}\right)-W_0^{(1)}(r)W_0^{(2)}(r)\right]dr\nonumber\\
&=&\frac{1}{2\varsigma^3}\bigg[\left(\frac{\mathcal{A}\alpha_0^8}{1512\lambda_D^{12}}-\frac{5\mathcal{A}\alpha_0^4}{12\lambda_D^8}+\frac{5\mathcal{A}}{\lambda_D^4}\right)+\frac{27\mu^2}{4\hbar^4\varsigma^4}\left(F-\frac{\mathcal{A}\alpha_0^6}{180\lambda_D^8}+\frac{\mathcal{A}\alpha_0^2}{\lambda_D ^4}\right)^2\nonumber\\
&&-\frac{9\mu\mathcal{A}}{2\hbar^2\varsigma^2}\left(F-\frac{\mathcal{A}\alpha_0^6}{180\lambda_D^8}+\frac{\mathcal{A}\alpha_0^2}{\lambda_D ^4}\right)\left(-\frac{\alpha_0^8}{4620 \lambda_D^{11}}+\frac{\alpha_0^6}{135\lambda_D^9}+\frac{\alpha_0^4}{7\lambda_D^7}-\frac{6\alpha_0^2}{5 \lambda_D^5}-\frac{2}{\lambda_D^3}\right)\bigg].\label{EQ20e}\nonumber\\
\end{eqnarray}
\end{subequations}
With Eqs. (\ref{EQ20a}-\ref{EQ20e}), we obtain the approximate energy eigenvalues  and the wavefunction of the hydrogen atom in laser-plasma encircle by electriced field as:
\begin{eqnarray}
E_{KH}\approx E_{KH}^{(0)}+\left(\frac{\mathcal{A}\alpha_0^8}{11340\lambda_D^9}+\frac{\mathcal{A}\alpha_0^6}{315\lambda_D ^7}-\frac{\mathcal{A}\alpha_0^4}{15\lambda_D ^5}-\frac{2\mathcal{A}\alpha_0^2}{3\lambda_D ^3}+\frac{2\mathcal{A}}{\lambda_D}\right)+E_{KH}^{(1)}+E_{KH}^{(2)}+E_{KH}^{(3)}+....,
\label{EQ21}
\end{eqnarray}
and
\begin{equation}
\Psi_0^{E_{_{KH}}}(\bm{r})\approx2\varsigma^{3/2}r\exp{\left(-\varsigma r\right)}\exp{\left(-\sqrt{\frac{2\mu}{\hbar^2}}\int_{}^r\left(W_{0}^{(1)}(\varrho)+W_{0}^{(2)}(\varrho)\right)d\varrho\right)},
\end{equation}
respectively. The behaviour of energy eigenvalues  of hydrogen atom in quantum plasma as it interacts with electric field and exposed to linearly polarized intense laser field radiation as a function of various model parameters has been shown in Table 1 and Figure {\ref{fig2}}. As it can be seen in Table 1, increasing the intensity of external electric field leads to corresponding increment in the bound state energy of hydrogen atom and the energy spacing becomes proliferated for a fixed Debye screening length. However, for a fixed weak external electric field, the scenario is quite different. The energy eigenvalues dwindle with increasing $\lambda_D$ and the energy spacing increases. The results in this table show that for a very weak external electric field an huge Debye screening length, the energy levels become more negative and the system becomes strongly repulsive.
\begin{table}[h!]
{\scriptsize
\caption{Energy eigenvalues (in a.u.) of hydrogen atom in quantum plasma as it interact with electric field and expose to linearly polarized intense laser field radiation.} \vspace*{10pt}{
\begin{tabular}{ccccccccccccccccc}\hline
{}{}&{}&{}&{}&{}&{}&{}&{}&{}&{}&{}&{}&{}&{}&{}&{}&{}\\[1.0ex]
$F$&&0.0001&&0.0004&&0.001&&0.004&&0.01&&0.04\\[3.5ex]\hline\\
$E_{KH}^{(\lambda_D=100)}$&&-1.9799255&&-1.9797005&&-1.9792506&&-1.9770016&&-1.9725072&&-1.9501083\\[3.5ex]\hline\\
$\lambda_D$&&5&&10&&20&&40&&80&&100\\[3.5ex]\hline\\
$E_{KH}^{(F=0.01)}$&&-1.5959955&&-1.7929741&&-1.8925671&&-1.9425144&&-1.9675077&&-1.9725072\\[3.5ex]\hline\hline\end{tabular}}}
\label{tab1}
\end{table}
To substantiate the results in Table 1, in Figure 2(a), we	scrutinize the behaviour of eigenvalues  of hydrogen atom as a function of laser-dressing parameter. For a particular external electric field, the energy shift is $0$ for $\alpha_0<0.06$ with $\lambda_D=1$. But as $\alpha_0$  proliferates, the energy level diminishes monotonically and becomes more negative. A significant change is seen in the localization of the bound state. However, when the Debye screening length is increased to $4$ as we have in figure 2(b), a slight shift is seen in the bound state energies with a pronounced characteristics even when the intensity of the external electric field changes. Figures 2(a) and 2(b) show the susceptibility of eigenvalues of hydrogen atom to $\lambda_D$. In fact, for the system to response to variation in $\alpha_0$ with various changes in electric field, a miniature Debye screening parameter must be deemed. 

Furthermore, in Figure 2(c), we study the behavior of eigenvalues of hydrogen atom as function of external electric field for various length of Debye screening parameter. For various $\lambda_D$, the system becomes strongly attractive as the intensity of electric field increases. In fact, this figure corroborates the result of Figure 2(b). For a strong $F$ and lilliputian Debye screening length, the energy levels tend towards positivity and the system becomes strongly attractive. In Figure 2(d), we elucidate Figure 2(c) further. As it can be seen, the energy levels under the influence of external electric field decrease monotonically with increase in Debye screening length until $\lambda_D\approx 25$. For $\lambda_D> 25$, no distinct variation can be discerned irrespective of variation in intensity of external electric field. In general, our numerical results evince that the repulsiveness of the system can be permuted via manipulation of external electric field. For instance, weak external electric field and a gargantuan length of Debye screening parameter, the system is strongly repulsive whereas for strong external electric field and small length of Debye screening parameter, the system is very attractive.

\section{Concluding Remarks}
We scrutinize the behavior of hydrogen atom's eigenspectra in a quantum plasma as it interacts with electric field and exposed to linearly polarized intense laser field radiation. Using the Kramers-Henneberger (KH) unitary translation, which is semiclassical counterpart of the Block-Nordsieck transformation in the quantized field formalism, the squared vector potential that appears in the equation of motion is eliminated and the resultant equation is represented in KH frame. Within this frame, the resulting potential and the corresponding wavefunction have been expanded in Fourier series and using Ehlotzky's approximation, we obtain a laser-dressed potential to simulate intense laser field. By fitting a more general exponential screened Coulomb potential into the laser-dressed potential, and then expand in Taylor series up to $\mathcal{O}(r^4,\alpha_0^9)$, we obtain the eigenspectra of hydrogen atom in laser-plasma encircle by electric field, within the framework of perturbation theory formalism. We have greatly simplified all mathematical expressions to the least possible terms so as to ensure this letter will not be only readable to the experts but to as well be understand by graduate students. We hope that, this study will inspire furtherance in future by exploring the molecular system in laser-plasma and also studying equation (\ref{EQ10}) for complex states (i.e. $n\ge1$).

\section*{Acknowledgments}
We thank the referees for their positive and enlightening comments and suggestions, which have greatly helped us in making improvements to this Letter. In addition, B.J.F. acknowledges eJDS and Dr. Oluwasesan Adeniran Falaiye. This work was partially supported by 20160978-SIP-IPN, Mexico.

\end{document}